\documentclass[a4paper]{aa} 
\usepackage[dvips]{graphics}
\usepackage{graphicx}
\usepackage{xspace}
\usepackage{amssymb,amsmath} 
\newcommand{\ascc}{\mbox{ASCC-2.5}\xspace}
\newcommand{\clucat} {\mbox{COCD}\xspace}
\newcommand{\starcat}{\mbox{CSOCA}\xspace}
\newcommand{\dias}{\mbox{DLAM}\xspace}
\renewcommand{\ni}{\noindent}

\usepackage{txfonts}
\begin{document}

\titlerunning{109 new Galactic open clusters}
\authorrunning{Kharchenko et al.}

\title{109 new Galactic open clusters\thanks{In fact, during
this search of new clusters in the catalogue
ASCC-2.5, we discovered 130 clusters. It turned out that 21 of them
are listed in the online list DLAM as private communications.
We stress the point
that this paper is the first presentation of these 21 clusters
in a refereed publication. Table~\ref{new_tab} including only
coordinates and sizes of the 109 new clusters as well as
Table~\ref{conf_tab} with coordinates and sizes of
the 21 confirmed clusters are available
only in the electronic version of the Journal at
http://www.edpsciences.org/aa. The complete set of data files for all
130 clusters is only available in electronic form at the CDS
via anonymous ftp to cdsarc.u-strasbg.fr (130.79.128.5)
or via http://cdsweb.u-strasbg.fr/cgi-bin/qcat?J/A+A/}}

\author{N.V.~Kharchenko \inst{1,2,3} \and
	     A.E.~Piskunov	\inst{1,2,4} \and
        S.~R\"{o}ser    \inst{2},
     	  E.~Schilbach	\inst{2} \and
     	  R.-D.~Scholz	\inst{1}
       }

\offprints{R.-D.~Scholz}

\institute{Astrophysikalisches Institut Potsdam, An der Sternwarte 16,\\
             D--14482 Potsdam, Germany\\
             email: nkharchenko@aip.de, apiskunov@aip.de, rdscholz@aip.de
         \and
          Astronomisches Rechen-Institut, M\"onchhofstra{\ss}e 12-14,
             D--69120 Heidelberg, Germany\\
             email: nkhar@ari.uni-heidelberg.de, piskunov@ari.uni-heidelberg.de,
                    roeser@ari.uni-heidelberg.de, elena@ari.uni-heidelberg.de
         \and
         Main Astronomical Observatory, 27 Academica Zabolotnogo Str.,
             03680  Kiev, Ukraine\\
             email: nkhar@mao.kiev.ua
           \and
           Institute of Astronomy of the Russian Acad. Sci.,
	   48 Pyatnitskaya Str., Moscow 109017, Russia\\
	   email: piskunov@inasan.rssi.ru
	  }

\date{Received ... December 2004; accepted ...}

\abstract{ \ni  We present a list of 130 Galactic Open Clusters,
found in the All-Sky Compiled Catalogue
of 2.5 Million Stars (\ascc).
For these clusters we determined and publish a homogeneous set of
astrophysical parameters such as size, membership, motion, distance
and age.
In a previous work
520 already known open clusters out of the sample of 1700 clusters from
the literature were confirmed in the \ascc using
independent, objective methods.
Using these methods the whole sky was systematically screened
for the search of new clusters.
The newly detected clusters show the same distribution over the sky
as the known ones. It is found, that without the {\em a-priori}
knowledge about existing clusters our search lead to clusters which are,
{\em on average}, brighter, have more members and cover larger angular radii
than the 520 previously known ones.
}


\maketitle

\section{Introduction}

For many years the major sources of open cluster lists were based on visual
inspection of photographic plates.  The present-day highly homogeneous and
accurate all-sky surveys like the Hipparcos and Tycho catalogues
(ESA~\cite{hip}), or the 2MASS
near-IR survey (Cutri et al.~\cite{2mas}) gave new impetus to a
systematic search of new clusters. Platais, Kozhurina-Platais \&
van Leeuwen (\cite{pla98}) profited from
the use of Hipparcos proper motions and parallaxes and detected six nearby
associations and nine candidates of open clusters. Using  
photometric and kinematical
data of the Tycho-2 catalogue (H{\o}g et al.~\cite{tyc2}), 
Alessi, Moitinho \& Dias (\cite{aea03}) 
detected 11 new clusters and determined their ages, geometric and kinematical
parameters. 
Dutra et al.~(\cite{dea03})
and Bica et al.~(\cite{bea03}) searched the 2MASS for compact embedded clusters
in the direction of known nebulae. The visual inspection of $J,H,K_s$ images has
lead to the discovery of 346 infrared clusters, stellar groups and candidates all over
the Milky Way.
All new optical clusters and candidates are listed in a catalogue
by Dias et al.~(\cite{daml02}). These authors also maintain
an online list of catalogues (\dias hereafter)
\footnote{\texttt{http://www.astro.iag.usp.br/\~{}wilton/}},
which is updated in regular intervals.

Our work is based on a catalogue of 2.5 million stars with proper motions in
the Hipparcos system and $B$, $V$ magnitudes in the Johnson photometric
system,  spectral types (\ascc; Kharchenko~\cite{kha01}) and radial velocities,
if available in the Catalogue of radial velocities of galactic stars with high
precision astrometric data (Kharchenko, Piskunov \& Scholz \cite{crvad}). 
The \ascc can be retrieved from the 
CDS\footnote{ftp://cdsarc.u-strasbg.fr/pub/cats/I/280A}; a detailed 
description of the catalogue can be found in Kharchenko~(\cite{kha01})
or in the corresponding ReadMe file at the CDS.
In a previous paper (Kharchenko et al.~\cite{starcat}, referred hereafter as
Paper~I), we used the \ascc to identify known open clusters and compact
associations, and developed an iterative pipeline for the construction of
cluster membership based on combined spatial/kinematical/photometric  criteria.
For 520 known clusters a uniform set of structural (location, size), kinematical
(proper motions and radial velocities) and evolutionary (age)  parameters was
derived (Kharchenko et al.~\cite{clucat}, referred hereafter as Paper~II).  The
results encouraged us to start a search for new clusters in the \ascc.

\begin{figure*}[t]\resizebox{\hsize}{!}
{\includegraphics[bbllx=60,bblly=195,bburx=525,bbury=385]
{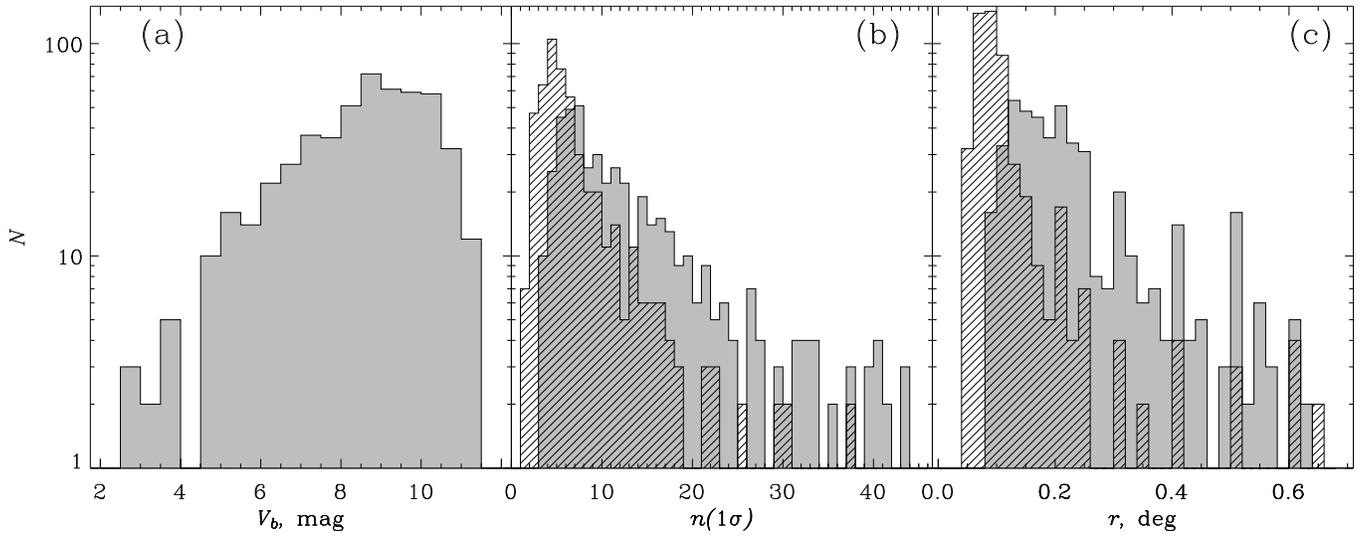}}
\caption{Distribution of the \clucat clusters over magnitude of the brightest
star among the most probable cluster members (a), over number of most probable 
members in a cluster (b), over angular radius of a cluster (c). The hatched
histograms in (b) and (c) are for core radii, whereas the filled histograms are
related to corona radii. For convenience, long tails in the distributions in
(b) and (c) are truncated.
}\label{dnall_fig}
\end{figure*}
\suppressfloats

The present paper describes this systematic search of new open clusters.
Instead of a visual inspection of sky surveys, we implement a multi-factor
search pipeline, which is based on the analysis of properties of known clusters
already identified in the \ascc. As a result we could increase our sample of
clusters present in the \ascc by about 20\%, determine memberships and derive a
uniform set of basic astrophysical parameters in the same way as for the 520
previously known clusters.

The paper has the following structure. In Sec.~\ref{known_sec} we discuss the
properties of known clusters identified in the \ascc. These properties give us
useful hints for the search of new clusters. In Sec.~\ref{search_sec} we
present details of the search procedure applied. In Sec.~\ref{results_sec} we
describe the sample of the newly discovered clusters and compare their
properties with those of known clusters already identified in the \ascc. In
Sec.~\ref{concl_sec} we summarise the results.

\section{Properties of known open clusters identified in the
ASCC-2.5}\label{known_sec}

For each star, the \ascc gives the equatorial coordinates, proper motions,
$B$, and  $V$ magnitudes. 
Only for a minority of them 
spectral and luminosity classes, and radial velocities are also known.
Therefore, starting from the data content of the \ascc, we suggest and adopt 
the following strategy of
searching new open clusters. This strategy is based on a clustering analysis 
in the multi-dimensional space of equatorial coordinates and proper motions 
with a follow-up check of colour-magnitude distributions of the candidates.
Since the completeness and especially the accuracy of the \ascc data show
a strong dependence on stellar magnitude, a straightforward search routine
must take these correlations into account. Furthermore, a successful approach
requires a set of starting parameters which are related to typical
properties of the given survey (e.g. mean surface density of stars, the 
limiting magnitude, wavelength range). The choice of starting parameters 
should  yield 
a reasonable relation between the number of cluster candidates selected at
the beginning and the number of real clusters confirmed at the end.

In this work we make use of the experience we got from the identification
of known open clusters in the \ascc. In that study we could find 520 of
some 1700 known clusters (Paper~I), 
and we derived cluster parameters such as sizes,
distances, ages, and space velocities
(Paper~II). The resulting parameters and supplementing
information on these clusters were gathered in the Catalogue of Open Cluster
Data (\clucat) supplemented by the Open Cluster Diagrams Atlas (OCDA). 
The main reason
that ``only'' 30\% of known open clusters were confirmed with the \ascc data
is the relatively bright limiting magnitude ($V_{lim} \approx 12$) of the 
catalogue. Also, a number of clusters from the 1700 known ones did not pass 
at least one of the criteria based on spatial, kinematical or 
photometric ($B$, $V$) data.  Considering the distribution of appropriate 
parameters of the confirmed clusters,
we can define selection criteria which will help to improve the chance of
detecting new clusters in the \ascc. 
This approach also gives a statistical basis for the strategy of searching.

The evident selection parameters among the number of cluster properties are
those related to the population and structure of a cluster as an enhanced 
density of brighter stars following a cluster main sequence. 
These properties can be translated to 
lower level descriptors like the number of bright cluster members located
within a specified area around the cluster centre.

In Fig.~\ref{dnall_fig} the distributions of 520 clusters from the \clucat 
(see Paper~II) are shown as functions of three relevant cluster 
parameters: the magnitude of the brightest star among the most 
probable cluster members, the number of the
most probable members in a cluster, and the angular size of a cluster. 
The most probable members are
defined in Paper~I as stars of which the individual proper motions, magnitudes, and 
colours deviate from the mean proper motion and the ``isochrone'' of
the cluster by less than one $\sigma$ $rms$
(``1$\sigma$-members'' i.e., stars with  membership probabilities $P \ge 61$\%).
From Fig.~\ref{dnall_fig}  we may conclude that a typical 
cluster from the \clucat has more than seven 1$\sigma$-members with at 
least one star brighter than $V$ = 9. Inner cluster regions have  a higher
stellar surface density, they are called cores in Paper~II and 
include typically five $1\sigma$-members. 
Although for a few clusters in the \clucat, the core and corona radii reach  
sizes of 3 and 6 degrees, respectively, for the vast majority (70-80\%) of the
clusters the members are concentrated within 0.15 deg (i.e. a typical core 
radius), but some members are found up to 0.30 deg (i.e. a typical corona 
radius) from the cluster centre.

\begin{table}[t]
\caption{Threshold values of clustering descriptors adopted in this study}
\label{parm_tab}
\setlength{\tabcolsep}{6pt}
\begin{tabular}{lccccc}
\hline
\rule{0mm}{3mm}
Descriptor &$\hat{V}_s$   & $\hat{r}_s$   & $\hat{r}_c$  & $\hat{n}_s$ & $\hat{n}_c$ \\
Value      &9.5 mag & 0.3 deg &0.15 deg&  8    &  5    \\
\hline
\end{tabular}\\
\end{table}

\section{Description of the search procedure}~\label{search_sec}

Our search procedure begins with the selection of ``seeds'', i.e. 
bright stars with magnitudes $V \le \hat{V}_s$ where $\hat{V}_s$ is the 
adopted 
limiting magnitude. At this step, we consider each seed as it were the most
probable kinematical member in the centre of a cluster. Then, we select all \ascc stars located 
at distances less than the adopted cluster radius $\hat{r}_s$. The derived 
sample is checked for a clustering in the multi-dimensional space of 
coordinates and proper motions. If a clustering could be revealed,
we apply our general pipeline of selecting cluster members in a celestial 
field and determine cluster parameters as described in Papers~I and II.
The pipeline is
run twice. For candidates of clusters, the first run provides initial 
estimates of geometric parameters and distances from the Sun. Also, it
removes those candidates for which we are
not able to derive these parameters (e.g. due to false clustering, or a lack of 
necessary data). The second run removes co-moving (non-member) stars and 
provides the final 
list of cluster members as well as the complete set of cluster parameters.

The crucial point of the search strategy is the selection of the threshold 
parameters which provide optimum starting conditions for a decision whether or 
not a real clustering exists. A common proper motion differing significantly
from the field would be a good criterion. But in general, we should assume 
that unknown clusters would have relatively small proper motions
(otherwise, they would already have been found). By increasing $\hat{V}_s$ and 
$\hat{r}_s$, we would find more cluster candidates, but the number 
of clusters really confirmed at the end would grow slowly and finally stop.
From preliminary tests we found that quite a reasonable
``cost-to-performance relation'' can indeed be achieved with threshold 
parameters, which are based on the statistics given in Sec.~\ref{known_sec}.

The search procedure uses the descriptors and their thresholds
as listed in Table~\ref{parm_tab}. 
The quantity $\hat{r}_s$ is the maximum search radius (analogue to
the cluster radius). In order to take into account the expected 
negative gradient of stellar
density in a cluster, we introduce a core radius $\hat{r}_c$. The minimum numbers
of members  within the cluster area and core are called $\hat{n}_s$ and
$\hat{n}_c$, respectively. The proper motions of members  
must follow the proper motion of the 
corresponding seed (i.e. analogue to $1\sigma$-members). 

The detailed procedure for searching new clusters consists of the following 
steps:

\begin{enumerate}

\item\textit{Selection of seed stars}. All \ascc stars with $V<\hat{V}_s$
      and $B-V<2$ mag which are not $1\sigma$-members of known clusters
      are considered as seeds. Altogether, about 221~000 stars have been 
      selected for further tests.

\item\textit{Construction of cluster candidates}. In a circle centred at a seed
      with radius $\hat{r}_s$, we select all \ascc stars 
      which proper motions are known with a mean error 
      $\varepsilon_{pm}<10$~mas/yr. Following the definitions in Paper~I and
      assuming the corresponding seed star $s$ to be
      a cluster member and its proper motion $\mu^s_x$, $\mu^s_y$ to represent the 
      mean cluster proper motion, we compute the kinematical probability $P^s_i$ 
      of the cluster membership for each star $i$ in the circle as
            \[
       P^s_i =  \exp \left\{-\frac{1}{4}\left[
                 \left(\frac{\mu_x^i - \mu^s_x}
                               {\varepsilon_{\mu^i_x}+\delta\varepsilon}\right)^2 +
                 \left(\frac{\mu_y^i - \mu^s_y}
                               {\varepsilon_{\mu^i_y}+\delta\varepsilon}\right)^2
                 \right]\right\},
       \]

      here $\mu_x$ and $\mu_y$ correspond to $\mu_\alpha\cos\delta$ and 
      $\mu_\delta$, respectively. $\delta\varepsilon$ is a correction for 
      external error of the proper motions in the \ascc (see Paper~I). 
      A sub-sample of stars with $P^s_i > 61$\% (``$1\sigma$-members'') is
      separated. If this sub-sample consists of $n_s\geqslant\hat{n}_s$ stars 
      with at least $\hat{n}_c$ stars within the inner circle of a radius 
      $\hat{r}_c$, we include this area into the following tests.
       After this second step we have obtained 4767 sky fields containing 
       cluster candidates. Some of them are overlapping areas.

\item\textit{Elimination of overlapping areas}. If neighbouring cluster 
      candidates have several stars in common, only those with the largest
      number $\hat{n}_s$ have been considered. Cases of double and triple
      overlaps have been treated automatically, whereas a few cases of 
      quadruple overlapping have been handled manually. After this step only 
      2472 candidate areas were retained for the next step.

\begin{table*}[tb]
\caption{List of newly-discovered clusters (cluster coordinates $\alpha_c$ in hours, $\delta_c$ in degrees, respectively for J2000). The cluster radii
$r_{cl}$ (in degrees) are also given. Note that the complete set of cluster parameters
is available only in electronic form via the CDS (see Sec.~\ref{results_sec}).}
\label{new_tab}
\setlength{\tabcolsep}{6pt}
\begin{tabular}{lrrr|lrrr|lrrr}
\hline
\rule{0mm}{3mm}
Cluster &$\alpha_c$&$\delta_c$&$r_{cl}$&Cluster &$\alpha_c$&$\delta_c$&$r_{cl}$&Cluster &$\alpha_c$&$\delta_c$&$r_{cl}$\\
\hline

ASCC 1  &  0.160&   62.68& 0.20&ASCC 43 &  7.885&$-$28.17& 0.35&ASCC 85  & 16.792&$-$45.46& 0.08\\
ASCC 2  &  0.331&   55.71& 0.30&ASCC 45 &  8.264&$-$35.65& 0.20&ASCC 87  & 17.050&$-$28.45& 0.15\\
ASCC 3  &  0.519&   55.28& 0.21&ASCC 46 &  8.276&$-$48.51& 0.40&ASCC 88  & 17.113&$-$35.60& 0.30\\
ASCC 4  &  0.886&   61.58& 0.40&ASCC 48 &  8.575&$-$37.61& 0.32&ASCC 90  & 17.652&$-$34.80& 0.15\\
ASCC 5  &  0.966&   55.84& 0.13&ASCC 51 &  9.300&$-$69.69& 0.66&ASCC 91  & 17.815&$-$37.36& 0.10\\
ASCC 6  &  1.787&   57.73& 0.30&ASCC 52 &  9.466&$-$54.26& 0.27&ASCC 93  & 18.137&$-$22.26& 0.08\\
ASCC 7  &  1.982&   58.97& 0.25&ASCC 53 &  9.632&$-$59.55& 0.31&ASCC 94  & 18.260&$-$14.99& 0.10\\
ASCC 8  &  2.347&   59.61& 0.30&ASCC 54 &  9.746&$-$54.44& 0.22&ASCC 95  & 18.268&$-$25.71& 0.10\\
ASCC 9  &  2.782&   57.73& 0.17&ASCC 55 &  9.905&$-$57.08& 0.23&ASCC 98  & 18.710&$-$33.63& 0.20\\
ASCC 11 &  3.538&   44.84& 0.35&ASCC 56 & 10.137&$-$64.37& 0.35&ASCC 99  & 18.818&$-$18.73& 0.10\\
ASCC 12 &  4.832&   41.73& 0.25&ASCC 57 & 10.180&$-$66.68& 0.38&ASCC 100 & 19.027&   33.57& 0.14\\
ASCC 13 &  5.222&   44.58& 0.70&ASCC 58 & 10.252&$-$54.97& 0.40&ASCC 101 & 19.227&   36.33& 0.15\\
ASCC 14 &  5.342&   35.22& 0.22&ASCC 59 & 10.337&$-$57.65& 0.35&ASCC 102 & 19.414&   29.95& 0.08\\
ASCC 15 &  5.377&   36.55& 0.20&ASCC 60 & 10.552&$-$58.48& 0.12&ASCC 104 & 19.648&   18.69& 0.15\\
ASCC 16 &  5.410&    1.80& 0.62&ASCC 61 & 10.769&$-$56.86& 0.32&ASCC 105 & 19.696&   27.38& 0.20\\
ASCC 17 &  5.420&   30.17& 0.25&ASCC 62 & 10.848&$-$60.10& 0.28&ASCC 107 & 19.809&   21.96& 0.07\\
ASCC 18 &  5.436&    0.82& 0.62&ASCC 63 & 10.931&$-$60.41& 0.15&ASCC 108 & 19.897&   39.37& 0.08\\
ASCC 19 &  5.463& $-$1.98& 0.80&ASCC 64 & 11.051&$-$60.92& 0.18&ASCC 109 & 19.900&   34.58& 0.15\\
ASCC 20 &  5.479&    1.63& 0.75&ASCC 65 & 11.185&$-$61.12& 0.22&ASCC 110 & 20.050&   33.57& 0.10\\
ASCC 21 &  5.483&    3.65& 0.80&ASCC 66 & 11.227&$-$55.42& 0.30&ASCC 111 & 20.187&   37.45& 0.20\\
ASCC 23 &  6.339&   46.67& 0.36&ASCC 67 & 11.692&$-$61.02& 0.20&ASCC 113 & 21.200&   38.60& 0.12\\
ASCC 24 &  6.479& $-$7.02& 0.35&ASCC 69 & 12.110&$-$69.77& 0.40&ASCC 114 & 21.667&   53.97& 0.08\\
ASCC 25 &  6.759&   24.60& 0.21&ASCC 70 & 12.250&$-$64.43& 0.30&ASCC 115 & 21.948&   51.48& 0.08\\
ASCC 26 &  6.840&    7.25& 0.17&ASCC 71 & 12.345&$-$67.52& 0.41&ASCC 116 & 21.976&   54.49& 0.08\\
ASCC 27 &  6.898& $-$4.39& 0.20&ASCC 72 & 12.550&$-$60.95& 0.25&ASCC 117 & 22.083&   62.27& 0.11\\
ASCC 28 &  6.901& $-$0.17& 0.30&ASCC 73 & 12.610&$-$67.29& 0.40&ASCC 119 & 22.320&   46.90& 0.08\\
ASCC 29 &  6.905& $-$1.65& 0.22&ASCC 74 & 13.597&$-$58.81& 0.20&ASCC 120 & 22.510&   57.21& 0.10\\
ASCC 30 &  6.950& $-$6.21& 0.26&ASCC 75 & 13.786&$-$62.42& 0.17&ASCC 121 & 22.512&   54.90& 0.10\\
ASCC 31 &  7.015&    3.50& 0.17&ASCC 76 & 13.871&$-$66.40& 0.35&ASCC 122 & 22.554&   39.61& 0.25\\
ASCC 33 &  7.053&$-$25.05& 0.90&ASCC 77 & 14.180&$-$62.33& 0.32&ASCC 123 & 22.710&   54.26& 0.28\\
ASCC 34 &  7.175&    6.07& 0.30&ASCC 78 & 15.085&$-$68.39& 0.16&ASCC 125 & 22.938&   62.75& 0.30\\
ASCC 35 &  7.211&    2.12& 0.40&ASCC 79 & 15.320&$-$60.73& 0.52&ASCC 126 & 23.105&   51.05& 0.10\\
ASCC 36 &  7.242&$-$21.12& 0.18&ASCC 80 & 15.410&$-$60.14& 0.25&ASCC 127 & 23.140&   64.85& 0.18\\
ASCC 37 &  7.301&$-$24.48& 0.16&ASCC 81 & 15.782&$-$50.98& 0.26&ASCC 128 & 23.343&   54.60& 0.15\\
ASCC 38 &  7.453& $-$5.55& 0.26&ASCC 82 & 15.790&$-$64.41& 0.30&ASCC 130 & 23.880&   62.44& 0.08\\
ASCC 39 &  7.550&$-$22.95& 0.30&ASCC 83 & 15.837&$-$52.80& 0.21&         &       &         &     \\
ASCC 40 &  7.560&$-$13.76& 0.18&ASCC 84 & 15.915&$-$60.74& 0.25&         &       &         &      \\
\hline
\end{tabular}\\
\end{table*}

\item\textit{Preliminary selection of cluster members and determination of
      cluster parameters}. This step and the following step~5 are based on the 
      pipeline developed in Papers~I and II for  member selection and 
      determination of cluster parameters. Now we consider a larger sky region 
      of $2\times2$ square degrees around each 
      remaining seed star. Additionally to the nominal kinematical selection, we
      carry out a simplified photometric selection, which removes stars located
      below the Main Sequence. Thus, at this stage, we keep possibly some
      red field stars which, by chance, show the same proper motions as a 
      given cluster candidate. As expected, the vast majority of cluster
      candidates does not show any Main Sequence and was excluded from
      further considerations. We are also forced to remove those cluster 
      candidates which do not contain at least one ``kinematical and photometric
      member'' with known spectral classification. In the current study, the
      spectral classification is the only information given in the
      \ascc which can be used for deriving  estimates of distance and color 
      excess. In a few cases, significant parallaxes (from Hipparcos) were available
      for the brightest probable cluster members. These parallaxes were used to
      check the derived distances.

      After this step, our list includes 308 cluster candidates with  
      preliminary determined cluster memberships and with a number of 
      preliminary cluster parameters like the position of the cluster's centre, 
      distance, and average proper motion.

\item\textit{Final determination of membership and cluster parameters}.
      At this stage, the standard pipeline is applied with complete kinematical 
      and photometric selection. After several iterations for each cluster, 
      co-moving red stars are excluded and the final cluster
      membership is established. The complete set of cluster parameters 
      including the cluster age is derived. If 
      cluster members are present in the \ascc sub-sample of stars with 
      radial velocities (Kharchenko, Piskunov \& Scholz \cite{crvad}), 
      the mean radial velocity of the cluster is computed.
      Only 130 clusters have passed this stage. The other 178 candidates have
      been excluded since after rejection of co-moving red stars, no more
      clustering in the space of coordinates and proper motions could
      be observed.
\item\textit{Visual inspection with sky maps}. 
      In a final step, supplementing the search for new clusters,
      we inspected Digitised Sky Survey 
      (DSS)\footnote{http://archive.stsci.edu/}  
      and SuperCOSMOS Sky Survey
      (SSS)\footnote{http://www-wfau.roe.ac.uk/sss/pixel.html} blue
      and red images. The size of an image around a given cluster was
      selected according to the determined cluster corona radius with
      a minimum size of $30\times30$~arcmin$^2$, and up to $180\times180$
      arcmin$^2$ for the largest clusters. The results of the visual
      inspection of the fields with new clusters are included in the notes
      to the extension of the OCDA. 
      For about 30 cluster areas the notes are with respect to the
      presence of nebulae and/or varying surface density of faint stars,
      which could not yet been mentioned in the \ascc sky maps.

\end{enumerate}

\section{Results of the search and comparison with properties of known
clusters}\label{results_sec}

We present the 130 new clusters in Tables~\ref{new_tab} and \ref{conf_tab}.
Among them we found 21 in the 15/feb/2004-update of \dias 
(see Table~\ref{conf_tab}). They
were not included in the COCD with its 520 clusters, because the COCD had already
been finished before this update became available. In \dias only cluster
centres and angular sizes are given for these 21 clusters, which were
privately communicated to the authors of the data base. We detected these clusters,
however, without using \dias data as preliminary input and determined
astrophysical parameters for them. Therefore, we consider them as
independently confirmed. The celestial positions of these 21 confirmed
clusters are given in Table~\ref{conf_tab}. On the other hand, about a dozen of the new
clusters (also privately communicated) in that update of \dias could not be
confirmed in our work.

\begin{figure*}[t!]
\resizebox{18cm}{!}
{\includegraphics[bbllx=20,bblly=30,bburx=565,bbury=790,angle=270,clip=]
{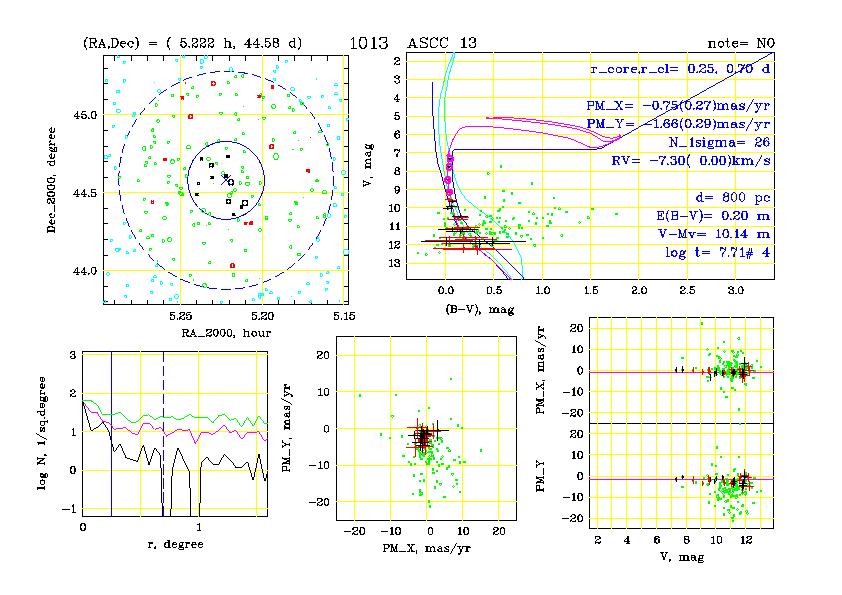}}
\caption{Example of the spatial, kinematical and evolutionary parameters of 
the new open cluster ASCC~13. Left upper panel is a sky map of the cluster
neighbourhood. The small circles are stars, their size indicates
stellar magnitude (only in this panel).
In all other diagrams stars are shown as grey dots.
The error bars indicate the $rms$-errors in the corresponding data for
$1\sigma$-members. The large circles in the sky map outline the cluster 
core (solid) and corona (dashed). The cross indicates the cluster center. 
The upper right panel is the CMD of the cluster. Bold circles show the stars 
used for the calculation of the average age of the
cluster. The curves are: the empirical ZAMS (leftmost), the red edge of the MS
(thick line to the right), the isochrone corresponding to the calculated age,
and the limits of the evolved MS (light curves). The legend within the CMD 
displays the derived parameters of the
cluster. The lower left panel shows radial profiles of the projected density.
The curves (from the top to the bottom) correspond to all stars, all members,
$1\sigma$-members. Vertical lines mark the core (solid) and cluster radii
(dashed), respectively. The middle panel is the
vector point diagram of proper motions.
The two rightmost panels are
``magnitude equation'' ($\mu_{x,y}-V$ relation) diagrams, showing the
proper motion of cluster members as a function of magnitude. The horizontal 
lines show the average proper motion of the cluster.
}\label{atl_fig}
\end{figure*}

\begin{figure*} 
\resizebox{18cm}{!}
{\includegraphics[bbllx=125,bblly=100,bburx=460,bbury=745,angle=90,clip=]
{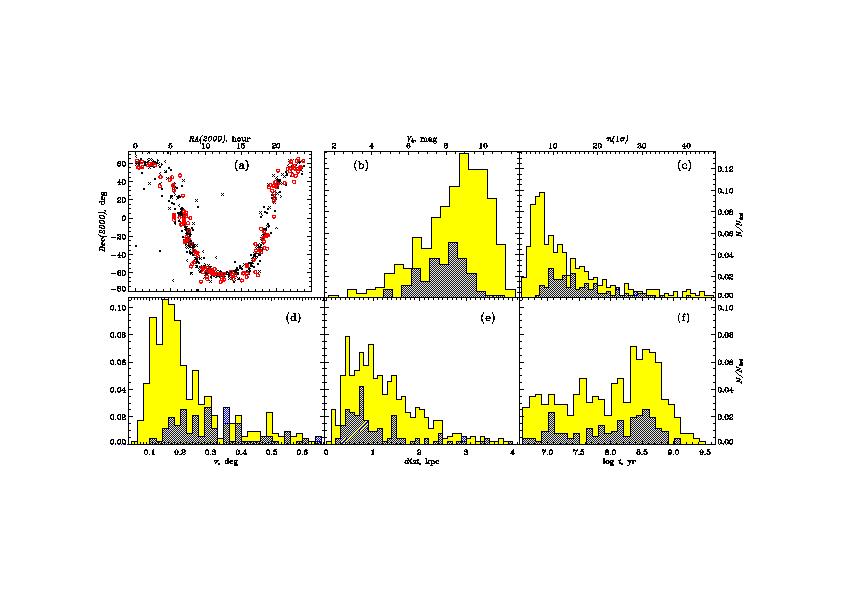}}
\caption{Open clusters identified with  \ascc data: the comparison of 
previously known (\clucat) and newly discovered
(\clucat Extension) clusters. Panel~(a) shows the distribution of clusters over 
the sky.  The crosses indicate known clusters, the circles are for newly
discovered ones. Histograms (b)-(f) are normalised to the number of known
clusters ($N_{tot}=520$), the filled and hatched histograms are for known and
newly discovered clusters, respectively. Panel~(b) is the distribution of
clusters over the magnitude $V_b$ of the brightest $1\sigma$-member. Panel~(c)
shows the distribution over the number of $1\sigma$-members in a cluster, 
whereas Panels~(d), (e), and (f) are the distributions of clusters over
angular radius, distance, and age, respectively. For convenience,
long tails in (c), (d) and (e) are truncated.
}\label{ncomp_fig}
\end{figure*}

At the moment of submission of this paper, we did not find any
published information on the other 109 clusters. Therefore, we consider them as
unknown clusters to date. They are listed in Table~\ref{new_tab}.

The current data refer to 130 sky areas and have the same content and format as
the \starcat   described in Paper~I. This file, which we call ``The 1st
Extension of \starcat'', includes 26\,778 stars, with 10\,161 of them located
within the determined cluster radii. According to the combined 
spatial/kinematical/photometric criteria, 6\,203 stars are classified as cluster
members and 2\,127 out of them as $1\sigma$-members. For all these  clusters, a
homogeneous set of basic astrophysical parameters is derived and the
corresponding cluster diagrams are prepared. Again, the data  are presented in
the same way as for the 520 known clusters (see Paper~II) and called ``The 1st
Extension of the \clucat'' and ``The 1st Extension of the OCDA'', respectively.
An example of a page in the Extension Atlas is shown in Fig.~\ref{atl_fig} 
for the open cluster ASCC~13. The complete set of data files, extending 
the \starcat, \clucat and OCDA, is available in electronic form only 
via the CDS\footnote{http://vizier.u-strasbg.fr/cgi-bin/VizieR}.

In Fig.~\ref{ncomp_fig} we compare two cluster samples retrieved from the \ascc.
One can see, that new clusters show the same distribution over the sky as the
already known objects (Fig.~\ref{ncomp_fig}a). Also, we may conclude that
the applied search strategy probably puts some limitation on the detection of
new clusters: the clusters from the \clucat Extension are, on average, brighter
(Fig.~\ref{ncomp_fig}b), they have more members (Fig.~\ref{ncomp_fig}c), and
they cover larger areas (Fig.~\ref{ncomp_fig}d) than the known clusters from 
the main \clucat catalogue. Whereas one new cluster is found at a distance
of 5 kpc, the bulk is located within 1~kpc from the Sun. The range of ages
is comparable for both samples, although a  higher fraction of young 
($\log t \approx 7$) clusters is observed in the \clucat Extension.

\section{Summary and outlook}\label{concl_sec}

Starting from 221,000 stars in \ascc which have passed the test for seeds
of possible open clusters, we applied our search procedure based on
spatial, kinematical and photometric criteria, and at the end of the
study, we found 130 new open clusters. For each of these clusters
a complete set of relevant parameters (memberships, locations, sizes, 
distances, ages, proper motions, and - for 69 clusters- radial velocities)
was derived. It seems to be a paradox, but we have now more basic
information on these new clusters than on many others reported as 
known clusters already for a long time. In our search for new clusters
we profited from all-sky astrometric and photometric surveys which became
available in recent years. In papers I and II we used the \ascc to identify known
open clusters, to re-define (or to confirm) the membership, and to derive 
a uniform set of astrophysical parameters for these clusters. The preliminary
information already published for these clusters was very helpful for
this study, too. In search for new
clusters, however, without any {\it a-priori} information, we needed at first
clear criteria to decide whether an apparent clustering was indeed a real
physical open cluster. 

\begin{table}[tb]
\caption{List of confirmed clusters 
(cluster coordinates $\alpha_c$ in hours, $\delta_c$ in degrees, 
respectively for J2000). Cluster radii $r_{cl}$ are in degrees.
Previous names and radii of the cluster candidates (i.e. the
only parameters which were provided by DLAM) are given
in brackets. Note that the full astrophysical parameter set determined in the
present paper is only available in electronic form via the CDS (see text).}
\label{conf_tab}
\setlength{\tabcolsep}{6pt}
\begin{tabular}{lrrc}
\hline
\rule{0mm}{3mm}
Cluster &$\alpha_c$&$\delta_c$&$r_{cl}$\\
\hline
ASCC 10  (Alessi-Teutsch 9) &  3.450&   35.04 &0.48 (0.36)\\
ASCC 22  (Ferrero 11)       &  6.242&    0.64 &0.18 (0.11)\\
ASCC 32  (Alessi 33)        &  7.033&$-$26.50 &0.50 (0.54)\\
ASCC 41  (Herschel 1)       &  7.784&    0.02 &0.36 (0.10)\\
ASCC 42  (Alessi-Teutsch 3) &  7.881&$-$53.01 &0.36 (0.29)\\
ASCC 44  (Alessi 34)        &  8.028&$-$50.57 &0.40 (0.48)\\
ASCC 47  (Alessi-Teutsch 7) &  8.530&$-$39.08 &0.50 (0.24)\\
ASCC 49  (Teutsch 38)       &  8.798&$-$37.99 &0.45 (0.37)\\
ASCC 50  (Alessi 43)        &  8.838&$-$41.72 &0.40 (0.38)\\
ASCC 68  (Alessi-Teutsch 8) & 12.049&$-$60.92 &0.20 (0.12)\\
ASCC 86  (Alessi J1701-58)  & 17.033&$-$59.01 &0.55 (0.30)\\
ASCC 89  (Alessi 24)        & 17.388&$-$62.64 &1.10 (0.75)\\
ASCC 92  (Alessi 31)        & 17.852&$-$11.88 &0.30 (0.22)\\
ASCC 96  (Ferrero 1)        & 18.335&$-$32.37 &0.25 (0.12)\\
ASCC 97  (Alessi 40)        & 18.616&$-$19.22 &0.43 (0.40)\\
ASCC 103 (Teutsch 35)       & 19.603&   35.67 &0.36 (0.18)\\
ASCC 106 (Alessi 44)        & 19.719&    1.60 &0.66 (0.50)\\
ASCC 112 (Alessi 46)        & 20.274&   52.10 &0.22 (0.20)\\
ASCC 118 (Alessi-Teutsch 5) & 22.140&   61.10 &0.21 (0.18)\\
ASCC 124 (Alessi 37)        & 22.802&   46.25 &0.30 (0.24)\\
ASCC 129 (Alessi J2327+55)  & 23.467&   55.60 &0.21 (0.21)\\
\hline
\end{tabular}\\
\end{table}

Comparing the histograms in Fig~\ref{ncomp_fig}(b,c) for the clusters from
the \clucat and from the \clucat Extension, we can conclude that the newly
discovered clusters are, on average, more prominent objects. There is
possibly a potential to find poorer clusters in the \ascc by diminishing
the threshold values in Table~\ref{parm_tab}. In this case, we must 
take into account a considerable increase of candidates with a lower
success rate for a final confirmation as undoubted clusters.

Assuming a similar distribution of $V_b$ and $n(1\sigma)$ for the known and
for the new clusters, we suspect from  (Fig~\ref{ncomp_fig}(b,c)) that the
\ascc should contain stars belonging to about 100 more open clusters still
unknown. But without accurate data at faint magnitudes, without any
knowledge of their distances, it will be difficult to confirm them.
We were already forced to reject a number of ``good'' candidates in step~4
of the search procedure due to the lack of spectral classification.
Therefore, a more successful approach has to wait till more accurate data at 
fainter magnitudes will become available.

\begin{acknowledgements}
This work was supported by the DFG grant 436~RUS~113/757/0-1, the RFBR
grant 03-02-04028, and by the FCNTP "Astronomy". We acknowledge the
use of the Simbad database  and the VizieR Catalogue Service operated
at the  CDS, France, and of the WEBDA facility at the Observatory of Geneva, 
Switzeland.
\end{acknowledgements}


\begin{thebibliography}{}
\bibitem[2003]{aea03}
        Alessi, B.S., Moitinho, A., \& Dias, W.S. 2003, A\&A, 410, 565
\bibitem[2003]{bea03}
         Bica, E., Dutra, C.M., Soares, J., \& Barbuy B.  2003,
         A\&A, 404, 223
\bibitem[2003]{2mas}
        Cutri, R. M. et al., 2003,
        in University of Massachusetts and Infrared Processing and Analysis Center,
        (IPAC/California Institute of Technology),
        VizieR On-line Data Catalog: II/246
\bibitem[2002]{daml02}
         Dias, W.S., Alessi, B.S., Moitinho, A.,\& L\'epine, J.R.D.,  2002,
         A\&A, 389, 871
\bibitem[2003]{dea03}
         Dutra, C.M., Bica, E., Soares, J., \& Barbuy B.  2003,
         A\&A, 400, 533
\bibitem[1997]{hip}
         ESA: 1997, Hipparcos and Tycho catalogues, ESA-SP 1200
         (Cat. I/239)
\bibitem[2000]{tyc2}
         H{\o}g, E., Fabricius, C., Makarov, V.V., et al.
         2000, The Tycho-2 Catalogue: Positions,
         proper motions and two-colour photometry of the 2.5 million
         brightest stars, Copenhagen, CD-ROM  distribution
\bibitem[2001]{kha01} 
         Kharchenko, N.V. 2001, Kinematics and Physics of Celestial 
         Bodies, 17, 409 
\bibitem[2004]{crvad} 
         Kharchenko, N.V., Piskunov, A.E., \& Scholz, R.-D. 2004, 
         Astron. Nachr., 325, 439
\bibitem[2004a]{starcat} 
         Kharchenko, N.V., Piskunov, A.E., R\"{o}ser, S., 
         Schilbach, E., \& Scholz, R.-D. 2004a, Astron. Nachr., 
         325, 743 (Paper~I)
\bibitem[2004b]{clucat}
         Kharchenko, N.V., Piskunov, A.E., R\"{o}ser, S.,
         Schilbach, E., \& Scholz, R.-D. 2004b, A\&A.,
         submitted (Paper~II)
\bibitem[1998]{pla98}
         Platais, I., Kozhurina-Platais, V., \& van Leeuwen, F. 1998,
         AJ, 116, 2423
         
\end{thebibliography}
\end{document}